\documentclass{ws-ijmpb}

\usepackage{psfrag}
\usepackage{epsfig}

\begin{document}

\markboth{S.T. Carr, J. Quintanilla, J. J. Betouras}
{Deconfinement and  quantum liquid crystalline States 
of dipolar fermions in optical lattices }

\catchline{}{}{}{}{}
%
%%%%%%%%%%%%%%%%%%%%%%%%%%%%%%%%%%%%%%%%%%%%%%%%%%%%%%%%%%%%%%%%%%%%

\title{DECONFINEMENT AND QUANTUM LIQUID CRYSTALLINE STATES 
OF DIPOLAR FERMIONS IN OPTICAL LATTICES  }

\author{Sam T. Carr $^1$, Jorge Quintanilla$^2$ and Joseph J. Betouras$^3$}

\address{$^1$School of Physics and Astronomy, University of Birmingham, \\
Birmingham B152TT, U.K.\\
sam.carr@physics.org\\
$^2$ISIS spallation facility, STFC Rutherford Appleton Laboratory, \\
 Harwell Science and Innovation Campus, OX11 0QX, U.K.\\
j.quintanilla@rl.ac.uk \\
$^3$University of St. Andrews \& SUPA, \\
North Haugh, St. Andrews KY16 9SS, U.K.\\
Joseph.Betouras@st-andrews.ac.uk}

\maketitle

\begin{history}
\received{Day Month Year}
\revised{Day Month Year}
%\accepted{(Day Month Year)}
%\comby{(xxxxxxxxxx)}
\end{history}

\begin{abstract}
We describe a simple model of fermions in quasi-one dimension that features interaction-induced deconfinement (a phase transition where the effective dimensionality of the system \emph{increases} as interactions are turned on) and which can be realised using dipolar fermions in an optical lattice \cite{2008-Quintanilla-Carr-Betouras}. The model provides a relisation of a "soft quantum matter" phase diagram of strongly-correlated fermions, featuring meta-nematic, smectic and crystalline states, in addition to the normal Fermi liquid. In this paper we review the model and discuss in detail the mechanism behind each of these transitions on the basis of bosonization and detailed analysis of the RPA susceptibility. 
\end{abstract}

\keywords{dipolar fermions; optical lattice; phases.}

\section{Introduction}

45 years have passed since John Hubbard identified {``understanding [...] the balance between bandlike and atomic-like behaviour''} as the key aim of research in strong correlations, and proposed the model that bears his name as the simplest embodiment of that conundrum \cite{1963-Gutzwiller,1963-Hubbard}. Yet, in spite of a plethora of systems lying somewhere between the Fermi liquid and Mott insulator or Wigner crystal states, the fundamental issue remains unresolved. In fact, even the simple Hubbard model has only been solved exactly in one \cite{1968-Lieb-Wu} and infinte \cite{1996-Georges-Kotliar-Krauth-Rozenberg} dimensions. Given the notable differences between the physics of these two extreme cases, one expects very rich behaviour in two and three dimensions - as suggested abundantly by experiment. 

Faced with the above difficulty, phenomenological scenarios have been put forward to deal with the wealth of experimental information on strongly-correlated quantum matter. For example, ``quantum liquid crystal'' phases with partially-broken symmetry (or, more generally, incomplete localisation) have been put forward as the missing links between the Fermi liquid and crystalline states of the fluid of electrons. \cite{1998-Kivelson-Fradkin-Emery} Another, closely related focus of attention have been dimensional crossovers \cite{2007-Giamarchi} and, in particular, collective phenomena changing the effective dimensionality of the system. (One advantage of focusing on the latter class of problems is that the known physics of the one-dimensional Hubbard model can be employed as a starting point, and the interaction introduced as a perturbation.)

In the latter class of phenomena we find the confinement hypothesis. It was originally
formulated \cite{1994-Clarke-Strong-Anderson} for an array of Luttinger liquids coupled
by a transverse single-particle hopping amplitude, $t_{\perp}$. The hypothesis states
that there is a finite, critical value of $t_{\perp}$ below which all coherent motion
becomes strictly one-dimensional. Indeed recent functional renormalization group
calculations \cite{2007-Ledowski-Kopietz} show that, for infinitesimally small
$t_{\perp},$ the ground state with a warped Fermi surface is unstable. The instability
can be described as an {}``ironing out'' of the Fermi surface, and is thus closely
related to other interaction-induced Fermi surface shape instabilities, notably the
Pomeranchuk instability \cite{1958-Pomeranchuk} leading into the highest-symmetry of the quantum liquid crystalline states: the nematic phase \cite{1998-Kivelson-Fradkin-Emery}. 

Another interesting question to ask is whether the \emph{opposite} of the confinement
transition is possible. By this we mean whether the almost flat Fermi surface of a system that is quasi-one-dimensional, in the sense that inter-chain hopping is very small, can
acquire some warping as a result of interactions. A similar phenomenon is known to occur in stacks of integer quantum Hall systems \cite{Betouras-Chalker,2005-Tomlinson-Caux-Chalker}. In this case, the chiral
one dimensional Luttinger liquids on the edges of the different layers couple together to
create a two-dimensional Fermi surface (the chiral Fermi liquid, for strong tunnelling
and interactions\cite{Betouras-Chalker} and for weak tunneling \cite{2005-Tomlinson-Caux-Chalker}). 

In this article we study this phenomenon in a different theoretical context, namely a
two-dimensional stack of chains, each of them containing free fermions. In the
absence of interactions the ground state of the system can be described as a
non-interacting Fermi gas with an almost flat Fermi surface. We introduce an
\emph{inter-chain} interaction, and address the stability of the quasi-one-dimensional
Fermi surface with respect to the perturbation. Interestingly, the inter-chain
interaction can lead to additional warping, and even the closing of the Fermi surface,
which thus becomes two-dimensional. In addition, we find that the model also exhibits a
crystalline state, for sufficiently strong interactions, which is entered through a
density wave instability. These meta-nematic and crystalline states compete with a third,
smectic phase corresponding to a different type of density wave instability. Thus the model realises some of the phenomenological phase diagram mentioned above \cite{1998-Kivelson-Fradkin-Emery}.

In the following sections we introduce the model, and discuss how it could be realised using cold atoms \cite{2008-Quintanilla-Carr-Betouras}. We will then describe in detail the mechanism by which the different phase transitions in the model come about, using a detailed analysis of the RPA susceptibility.
We also offer a discussion of some aspects of the physics of the system using arguments based on bosonization. 

\begin{figure}
\center{\includegraphics[width=0.35\columnwidth,keepaspectratio,angle=-90]{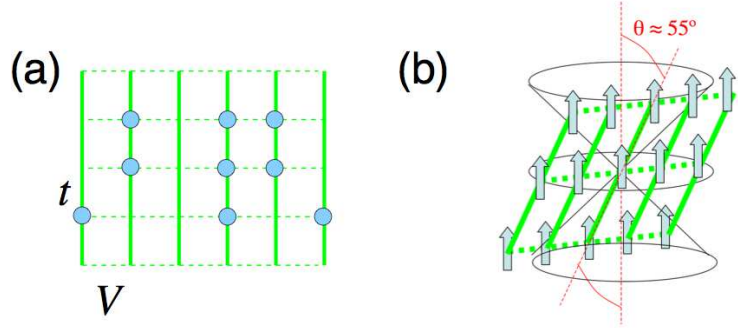}}
\caption{\label{fig:1}(a) Our Hamiltonian describes hopping mostly along a
set of
chains (with very little perpendicular hopping) plus a nearest-neighbour,
inter-chain interaction. (b) A two-dimensional,
strongly-anisotropic optical lattice with dipolar interactions can
provide an approximate experimental realistion (see text).}
\end{figure}

\section{\label{sec:Model}Model}

The unperturbed Hamiltonian is 
\begin{equation}
\hat{H}_{\mbox{hop}}=-t_{\|}\sum_{l,i}\hat{c}_{l,i}^{\dagger}\hat{c}_{l,i+1}-t_{\perp}\sum_{l,i}\hat{c}_{l,i}^{\dagger}\hat{c}_{l+1,i}+\mbox{H.c.}-\mu\sum_{l,i}\hat{c}_{l,i}^{\dagger}\hat{c}_{l,i},
\label{eq:H_hop}
\end{equation}
It describes a set
of chains kept at the same chemical potential, $\mu$. $\hat{c}_{l,i}^{\dagger}$ creates a fermion on the $i^{\mbox{th}}$
site of the $l^{\mbox{th}}$ chain. The particles
can hop easily \emph{along} the chains, with amplitude $t_{\|}$, but not so much \emph{between}
the chains, for which the hopping amplitude is $t_{\perp}\ll t_{\|}$. We assume the spin degree of freedom to be frozen (spinless,
or fully polarised, fermions -see below). 

The above Hamiltonian is almost diagonal
in the chain index, $l$, and therefore has a nearly-flat, quasi-1D Fermi surface.

The perturbation is an \emph{interaction} between particles sitting
on adjacent sites of \emph{different} chains:
\begin{equation}
\hat{H}_{\mbox{int}}=V\sum_{l,i}\hat{c}_{l,i}^{\dagger}\hat{c}_{l+1,i}^{\dagger}\hat{c}_{l+1,i}\hat{c}_{l,i}.
\label{eq:H_int}
\end{equation}
The full Hamiltonian, 
\begin{equation}
\hat{H}=\hat{H}_{\mbox{hop}}+\hat{H}_{\mbox{int}},
\label{eq:H}
\end{equation}
 is represented schematically in Fig.~\ref{fig:1}~(a). It describes
hopping along the chains (the solid lines in the figure), with only small inter-chain hopping, plus an inter-chain
interaction (represented by the dashed lines). 

\section{\label{sec:Experimental-Realisation}Experimental Realisation}

A new avenue into strong correlations has been opened recently in ultracold atomic gases \cite{2008-Bloch}. Indeed the Mott transition has been observed in very precise experimental relisations of the Hubbard model in a harmonic trap \cite{2008-Jordens-et-al,2008-Schneider-et-al}. In principle, a sequence of experiments on systems with different trap profiles could be used to deduce the phase diagram of the Hubbard model in the absence of the trap \cite{2007-Campo-Capelle-Quintanilla-Hooley}. Moreover, even simpler models can be realised to explore the physics of dimensional crossovers and phenomenological scenarios of strong correlations. Interestingly, our model can be realised, quite precisely, in an optical lattice setup loaded with quantum-degenerate dipolar fermionic molecules \cite{2008-Ni-et-al} or fermionic atoms with large magnetic dipoles such as $^{53}Cr$ \cite{foot:Cr}. 

The Hamiltonian (\ref{eq:H}) is theoretically convenient because
the unperturbed part features no interactions and can therefore be
described in terms of free fermions. Interactions are introduced perturbatively.
Moreover the interactions do not change the nature of each individual
chain, but take place between distinct chains exclusively. In this section we describe a possible realisation of this physics.

The atoms (or, equivalentely, molecules - though for concreteness we will assume atoms in what follows) are trapped in a
two-dimensional optical lattice by two pairs of counter-propagating
laser beams: see Fig.~\ref{fig:1}~(b). In the figure, the solid
and dashed lines represent their respective wavefronts. If the lasers
are sufficiently intense, the lattice can be described in the tight-binding
limit, with only one orbital per site. Moreover by making one of the
pairs of lasers much more intense than the other we can create {}``chains''
along which hopping can take place, but such that hopping between
different chains can be very small. These chains are represented by
the solid lines in Fig.~\ref{fig:1}~(b). The lattice sites are
where the solid and dashed lines cross. 

To suppress the interaction between atoms that are in the same chain, we exploit the
dependence of the dipole-dipole interaction, 
\begin{equation}
V\left(\mathbf{R}\right)=d^{2}\frac{1-3\cos^{2}\theta}{\left|\mathbf{R}\right|^{3}},
\label{dipole_interaction}
\end{equation}
on the angle $\theta$ between the vector giving the relative positions of the two dipoles,
$\mathbf{R}$, and the external field, $\mathbf{H}$. This interaction is identically zero
for $\theta=\arccos\left(1/\sqrt{3}\right)\approx54.736^{\mbox{o}}$. Thus aligning the
chains at this angle (as in the figure) we ensure that there are no interactions between
atoms in the same chain (on-site interaction, including the additional one due to the Van
der Waals forces between the atoms, is already suppressed by Pauli's exclusion principle,
for fully polarised fermions). 

Finally, by arranging the chains so that each site is on the horizontal from the closest
sites on the two adjacent chains we ensure that the interaction with those two sites is
maximally repulsive {[}$V\left(\mathbf{R}\right)=d^{2}/\left|\mathbf{R}\right|^{3},$
corresponding to $\theta=\pi/2$]. Longer-ranged inter-chain interactions with other sites
on the two adjacent chains  can be made comparatively weaker by making the lattice
constant very long along the longitudinal direction. This is aided by the relatively
rapid fall of the interaction potential with distance,
$\propto1/\left|\mathbf{R}\right|^{3}$. To achieve this in combination with keeping
hopping along the dotted lines to a minimum requires that the laser creating the solid
line wave fronts be very intense. This ensures a very high potential barrier for the
atoms to hop between chains. It is also desirable to orient the plane of the lattice so
as to ensure that all inter-chain interactions are repulsive. The details of how the precise realisation of our Hamiltonian can be achieved are given in Ref.~\cite{2008-Quintanilla-Carr-Betouras}.

\section{Meta-nematic phase transition} 

In order to establish which ground states may occur in this system we start by evaluating the stability of the Fermi surface shape.  The interaction in Hamiltonian (\ref{eq:H}) wants to avoid having two particles on the same site on neighboring chains.  Physically, this is very similar to interchain hopping -  if only one electron is present, it can lower its kinetic energy by hopping.  We may therefore expect the interaction to renormalize the effective interchain hopping to be larger than its bare value.  We investigate this possibility by using a restricted Hartree-Fock mean field theory similar to those used to study Pomeranchuk\cite{2004-Khavkine,2006-Quintanilla} and topological\cite{2006-Quintanilla} Fermi surface shape instabilities.

We use as a trial ground state a Slater determinant of plane waves,
 \begin{equation}
 \left|\Psi\right\rangle =\prod_{\mathbf{k}}\left[\left(1-N_{\mathbf{k}}\right)+N_{\mathbf{k}}\hat{c}_{\mathbf{k}}^{\dagger}\right]\left|0\right\rangle,
 \end{equation}
 determining the occupation numbers $N_{\bf k}=0,1$ by requiring that the momentum distribution minimizes $\langle \Psi \left| \hat{H} \right| \Psi \rangle$.
The momentum distribution $N_{\bf k}$ corresponds to a non-interacting Fermi gas with a renormalized dispersion relation
\(
  \epsilon_{\mathbf{k}}^*=-2t_{\|}\cos\left(k_{\|}\right)
  -2t_{\perp}^*\cos\left(k_{\perp}\right)-\mu^*.
  \label{eq:epskr}
\)
The structure of the interaction $\hat{H}_{\mbox{int}}$ is such that only the
perpendicular hopping is changed. It is given by the following self-consistency
equation:
\begin{eqnarray}
  t_{\perp}^{*} 
  & = & 
  t_{\perp}
  +
  \frac{V}{\Omega}\sum_{\mathbf{k}}\cos\left(k_{\perp}\right)N_{\mathbf{k}},
  \label{eq:tc*}
\end{eqnarray}
where $N_{\mathbf{k}} = \theta(-\epsilon_{\mathbf{k}}^*)$ is the Heaviside function. In what follows all references to the chemical potential will be to its renormalized value so we will omit the $*$
for that quantity.

\begin{figure}[t]
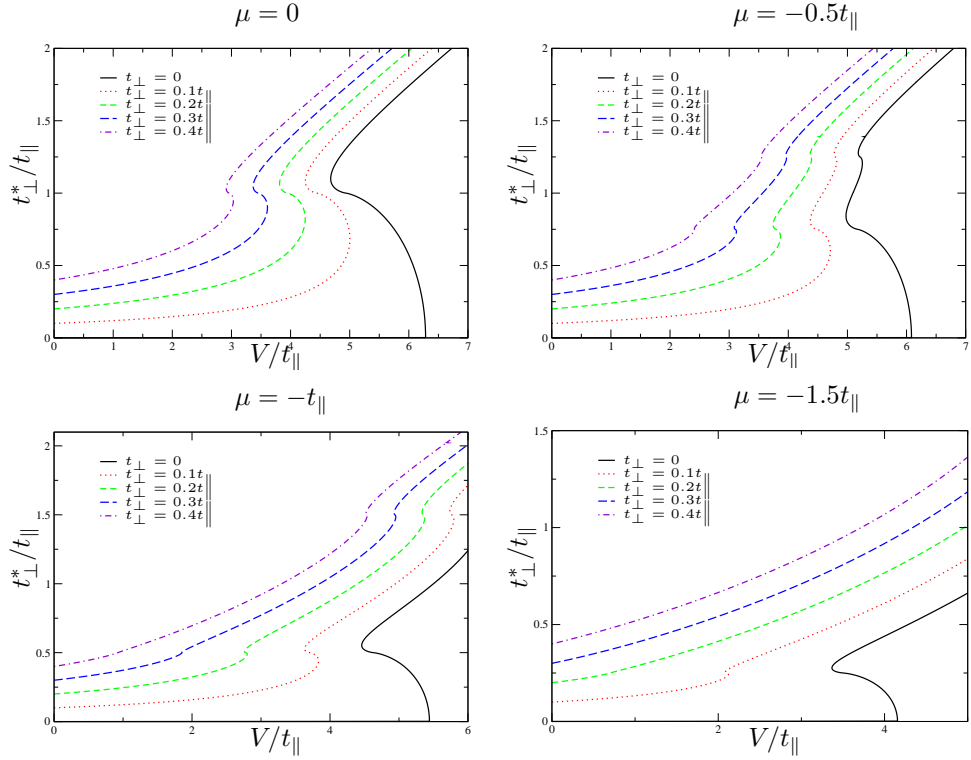

\begin{center}
\psfrag{lx_tp}{$t_\perp^*/t_\|$}
\psfrag{lx_V}{$V/t_\|$}
\psfrag{lx_m00}{$\mu=0$}
\psfrag{lx_m05}{$\mu=-0.5t_\|$}
\psfrag{lx_m10}{$\mu=-t_\|$}
\psfrag{lx_m15}{$\mu=-1.5t_\|$}
\psfrag{lx_f0}{{\tiny $t_\perp = 0$}}
\psfrag{lx_f1}{{\tiny $t_\perp=0.1 t_\|$}}
\psfrag{lx_f2}{{\tiny $t_\perp=0.2 t_\|$}}
\psfrag{lx_f3}{{\tiny $t_\perp=0.3 t_\|$}}
\psfrag{lx_f4}{{\tiny $t_\perp=0.4 t_\|$}}
\epsfig{file=mu.0.0.eps, width=6cm}\hspace{0.5cm}
\epsfig{file=mu.0.5.eps, width=6cm}\vspace{0.3cm}
\epsfig{file=mu.1.0.eps, width=6cm}\hspace{0.5cm}
\epsfig{file=mu.1.5.eps, width=6cm}
\end{center}
\caption{The renormalized hopping amplitude $t_\perp^*$ as a function of interaction strength $V$ for various different values of chemical potential $\mu$ and bare hopping $t_\perp$.   The values of $V$ where $t_\perp^*$  is not single-valued which ultimately corresponds to a jump in $t_\perp^*$ as $V$ is increased, mark the meta-nematic transition. A full discussion of these graphs is given in the main text.}\label{fig:mn1}
\end{figure}

\begin{figure}
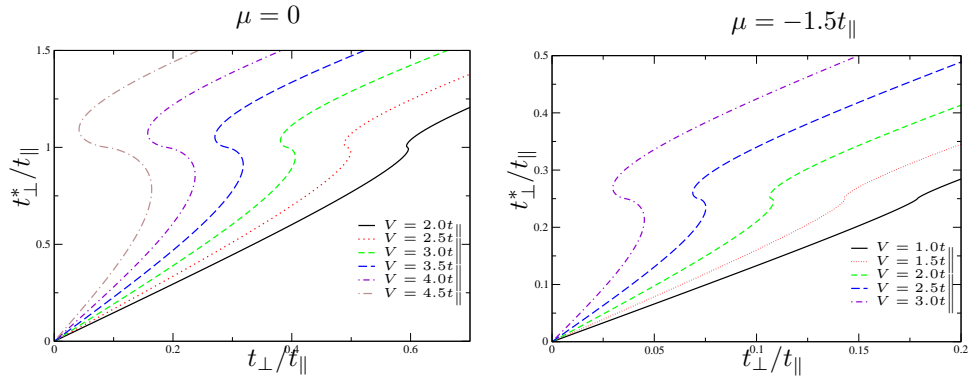

\begin{center}
\psfrag{lx_tpb}{$t_\perp/t_\|$}
\psfrag{lx_tp}{$t_\perp^*/t_\|$}
\psfrag{lx_m00}{$\mu=0$}
\psfrag{lx_m15}{$\mu=-1.5t_\|$}
\psfrag{lx_V10}{{\tiny $V=1.0 t_\|$}}
\psfrag{lx_V15}{{\tiny $V=1.5 t_\|$}}
\psfrag{lx_V20}{{\tiny $V=2.0 t_\|$}}
\psfrag{lx_V25}{{\tiny $V=2.5 t_\|$}}
\psfrag{lx_V30}{{\tiny $V=3.0 t_\|$}}
\psfrag{lx_V35}{{\tiny $V=3.5 t_\|$}}
\psfrag{lx_V40}{{\tiny $V=4.0 t_\|$}}
\psfrag{lx_V45}{{\tiny $V=4.5 t_\|$}}
\epsfig{file=tp_tpb.mu.0.0.eps, width=6cm}\hspace{0.5cm}
\epsfig{file=tp_tpb.mu.1.5.eps, width=6cm}
\end{center}
\caption{A different way of viewing the data: The renormalized hopping amplitude $t_\perp^*$ as a function of the bare hopping $t_\perp$ for various values of interaction strength $V$ and chemical potential $\mu$.}\label{fig:mn2}
\end{figure}

Numerical solutions to this equation are shown in Figs. \ref{fig:mn1} and \ref{fig:mn2}. As either the bare
inter-chain hopping $t_{\perp}$ or the interaction $V$ is increased, the renormalized hopping,
$t_{\perp}^*$, initially increases linearly but then has two bifurcation points,
between which lies a first-order jump to a higher value.  The order parameter of this phase
transition is the amount of delocalisation in the perpendicular direction,
$\psi \equiv \langle \hat{c}_{i,l}^{\dagger}\hat{c}_{i,l+1}+\mbox{H.c.}\rangle$. We refer to
the jump of $\psi$ as we vary $t_{\perp}$ as a meta-nematic transition in analogy with
meta-magnetism (where the magnetisation jumps under an applied magnetic field).

Focusing on Fig. \ref{fig:mn1}, we now discuss some features of this transition.  For fixed $\mu$, the transition always occurs at the same value of the renormalized hopping $t_\perp^*$, and in fact corresponds exactly to the point where the Fermi surface switches from open to closed, something we refer to as dimensional crossover from quasi-1D to quasi-2D.  We see also as we take the Fermi level from half-filling $\mu=0$ to nearer the bottom of the band $\mu=-1.5t_\|$ the transitions generally moving down to a lower energy.  In some of the lines, there are no transitions - which is when the bare perpendicular hopping $t_\perp$ is already large enough for the bare Fermi surface to be closed - and the renormalization occurs in a continuous manner.  Another point to note is that in some of the lines, there are two separate first order transitions - the second in fact corresponds to the value of $t_\perp^*$ where the Fermi-surface becomes open again, but in the other direction.  This gives strong hints that the meta-nematic transition is indeed a dimensional crossover phenomena - however this second transition requires very large renormalization of $t_\perp$ well outside any regime where Hartree-Fock is valid, and furthermore one would expect the electrons to crystallize before that point (see the next section) so we will not say anything further about the second transition.

\begin{figure}
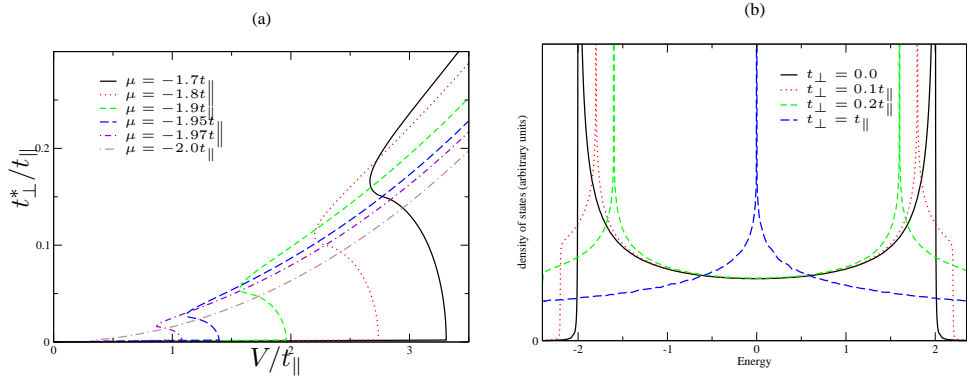

\begin{center}
\psfrag{lx_tp}{$t_\perp^*/t_\|$}
\psfrag{lx_V}{$V/t_\|$}
\psfrag{lx_m17}{{\tiny $\mu=-1.7 t_\|$}}
\psfrag{lx_m18}{{\tiny $\mu=-1.8 t_\|$}}
\psfrag{lx_m19}{{\tiny $\mu=-1.9 t_\|$}}
\psfrag{lx_m195}{{\tiny $\mu=-1.95 t_\|$}}
\psfrag{lx_m197}{{\tiny $\mu=-1.97 t_\|$}}
\psfrag{lx_m20}{{\tiny $\mu=-2.0 t_\|$}}
\psfrag{lx_tp00}{{\tiny $t_\perp = 0.0$}}
\psfrag{lx_tp01}{{\tiny $t_\perp = 0.1 t_\|$}}
\psfrag{lx_tp02}{{\tiny $t_\perp = 0.2 t_\|$}}
\psfrag{lx_tp10}{{\tiny $t_\perp =  t_\|$}}
\epsfig{file=lifshitz.eps, width=6cm}\hspace{0.5cm}
\epsfig{file=dos1.eps, width=6cm}
\end{center}
\caption{(a) A plot of the renormalized hopping $t_\perp^*$ as a function of interaction strength $V$ for different chemical potentials $\mu$ approaching the bottom of the band when the bare hopping $t_\perp=0$.  This figure clearly shows both the critical value of $V$ and the size of the first order jump decreasing towards zero as the bottom of the band is reached - i.e. the Lifshitz transition. (b) The density of states of the anisotropic 2D lattice for various values of the ratio between parallel and perpendicular hopping.  The van-Hove singularity in the density of states occurs at exactly the energy when the contour of constant energy just touches the edge of the Brioullin zone - i.e. when the Fermi surface switches from open to closed.  Enhanced scattering due to this singularity is the origin of the meta-nematic transition.The small tails at the band edges are a numerical artefact. The energy is measured in units of $t_{||}$.}\label{fig:lifshitz}
\end{figure}

As $V \to 0$, the meta-nematic transition becomes more and more weakly first order and requires a larger bare value of $t_{\perp}$. At strictly $V=0$ there is no longer a first-order phase transition, but the phenomenon survives at $t_{\perp}=t_{\|}+\mu/2$ as a `two-and-half' order Lifshitz transition\cite{lifshitz} - see Fig. \ref{fig:lifshitz}(a). On the other hand the transition is first order for any $V>0$. 

The meta-nematic transition results from enhanced scattering when the potential reaches the singularities at the edges of the 1D bands. This is a density of states effect - see Fig. \ref{fig:lifshitz}(b) and hence we expect it to be robust to quantum fluctuations present for large values of $V/t_{\|}$ and not taken into account by our mean field theory.

%\begin{figure}[bt]
%\centerline{\psfig{file=Fig_3.eps,width=3.65in}}
%\vspace*{8pt}
%\caption{Dependence of the renormalised transverse hopping $t_{\perp}^*$ on its bare value $t_{\perp}$ for $V=3t_{\|}$ and $\mu = -1.5
%t_{\|}$. The solid (dashed) lines correspond to solutions to the self-consistency
%equation (\ref{eq:tc*}) that minimize (maximize) the energy.Insets: momentum distribution in the first Brillouin
%zone  (a) just to the left of the bifurcation region and (b) just to the right, as indicated; and (c) dependence of the solutions on $V$ for $t_{\perp}=0$ (rightmost curve), $0.05t_{\|}$
%and $0.1t_{\|}$ (leftmost curve).}
%\end{figure}

\section{Crystallisation} 

The meta-nematic transition is not the only one possible in the system described by Eq.~(\ref{eq:H}).  In fact, for $t_\perp=0$, the Fermi surface is perfectly nested, leading to the possibility of a density wave instability at low temperatures - indeed, the strong coupling phase of Hamiltonian \ref{eq:H} is a crystalline state.  Another way of thinking of this is in terms of the interchain backscattering\cite{Klemm}.  Although in the cold atom set-up, the fermions are not charged, we will still refer to this as a charge density wave (CDW) instability in order to draw parallels with previous work on electronic models.

We probe the potential CDW instability by examining the Fourier transform of the dynamic susceptibility,
\begin{equation}
X(\mathbf{k},t)=i\left\langle \Psi \left| T \rho(\mathbf{k},t)\rho^{\dagger}(\mathbf{k},0)\right|\Psi\right\rangle.
\end{equation} 
Here $\rho(\mathbf{k})=\sum_{\bf q} c_{\bf q}^{\dagger}c_{{\bf q}-{\bf k}}$ is the Fourier transform of the local occupation number, and $\rho({\bf k},t)$ is its Heisenberg representation. 

The `noninteracting' susceptibility (Lindhard function) is given by
\begin{equation}
X_0 (\mathbf{k},\omega) = \int \frac{d^2 \mathbf{q}}{(2\pi)^2} \frac{N_{\mathbf{q}} - N_{\mathbf{q}+\mathbf{k}}}{\omega - \epsilon^*_\mathbf{q} + \epsilon^*_{\mathbf{q}+\mathbf{k}}}\label{eq:lindhard}.
\end{equation}
We note that this is written in terms of the renormalised dispersion relation $\epsilon_{\bf k}^{*}$ (technically the 'bubble' should be dressed).  We also point out that because the major contribution to the kinetic energy is perpendicular to the direction of the interaction, any vertex corrections to the bubble are small and may be neglected in lowest order.

The interaction may then be treated within the random phase approximation (RPA)
\begin{equation}
 X (\mathbf{k},\omega) = \frac{X_0 (\mathbf{k},\omega)}{1+2V\cos(k_{\perp}) X_0 (\mathbf{k},\omega)}.
 \end{equation}
An instability at wave vector $\mathbf{k}$ takes place if the static  component of the susceptibility diverges, 
  \(
    X(k_{\|},k_\perp,\omega=0) \to \infty.
  \) 

\begin{figure}
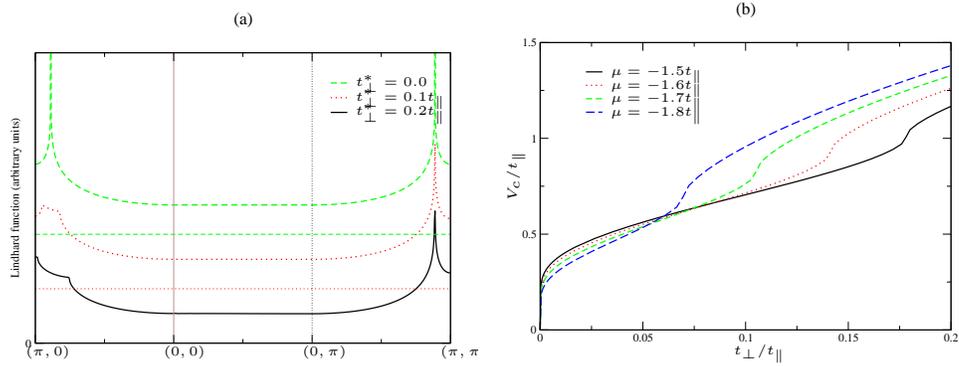

\begin{center}
\psfrag{lx_tp}{{\tiny $t_\perp/t_\|$}}
\psfrag{lx_Vc}{{\tiny $V_c/t_\|$}}
\psfrag{lx_m17}{{\tiny $\mu=-1.7 t_\|$}}
\psfrag{lx_m18}{{\tiny $\mu=-1.8 t_\|$}}
\psfrag{lx_m15}{{\tiny $\mu=-1.5 t_\|$}}
\psfrag{lx_m16}{{\tiny $\mu=-1.6 t_\|$}}
\psfrag{lx_tp0}{{\tiny $t_\perp^*=0.0$}}
\psfrag{lx_tp1}{{\tiny $t_\perp^*=0.1t_\|$}}
\psfrag{lx_tp2}{{\tiny $t_\perp^*=0.2 t_\|$}}
\psfrag{lx_1}{{\tiny $(\pi,0)$}}
\psfrag{lx_2}{{\tiny $(0,0)$}}
\psfrag{lx_3}{{\tiny $(0,\pi)$}}
\psfrag{lx_4}{{\tiny $(\pi,\pi)$}}
\epsfig{file=chi.eps, width=6cm}\hspace{0.5cm}
\epsfig{file=Vcrit_tbare.eps, width=6cm}
\caption{(a) The Lindhard function, Eq. \ref{eq:lindhard} as a function of $(k_\|,k_\perp)$ along the path $(\pi,0)\rightarrow (0,0) \rightarrow (0,\pi) \rightarrow (\pi,\pi)$.  For clarity, the axis has been shifted vertically for the three different values of $t_\perp$ plotted.  The singularity at $(2k_F,\pi)$ is clearly seen in the case when $t_\perp=0$, reducing to a sharp peak when $t_\perp$ is finite. (b) The critical $V_c$ from Eq. \ref{eq:rpa} plotted against the {\em bare} hopping $t_\perp$ for various values of $\mu$.  The humps are at the points when the Fermi surface switches from being open to closed - which results in much less nesting. }\label{fig:lindhard}
\end{center}
\end{figure}

\begin{figure}[bt]
\centerline{\psfig{file=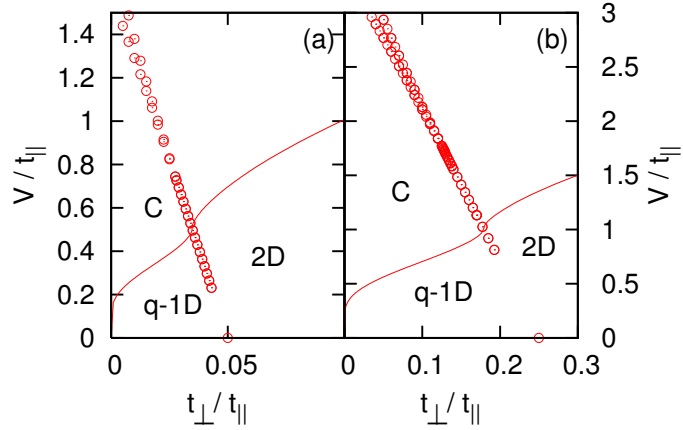,width=3.65in}}
\vspace*{8pt}
\caption{Phase diagram of the Hamiltonian of Eq.~(\ref{eq:H}) for $\mu = -1.9
t_{\|}$ (a) and $-1.5 t_{\|}$ (b). The circles track the two bifurcation
 of the first-order meta-nematic quantum phase transition between the quasi-one dimensional (q-1D) and two dimensional (2D) phases. The solid line marks the line of quantum critical points separating these two Fermi liquid states from the crystalline state (C).}\label{fig:phase_diagram}
\end{figure}

For $t_{\perp}^* \ll t_{\|}$ the Lindhard function is strongly peaked at $(2k_F,\pi)$, and in fact logarithmically divergent at this wavevector when $t_\perp^*=0$ - see Fig. \ref{fig:lindhard}(a). This is due to the strong nesting of the quasi one-dimensional Fermi surface, and it makes the system unstable to a CDW of that periodicity at a critical coupling $V$ given by the following Stoner criterion:
\begin{equation}
    1=2V_cX_0(2k_F,\pi,\omega=0),
   \label{eq:rpa}
\end{equation} 
which clearly gives $V_c=0$ in the perfectly nested Fermi-surface when $t_\perp=0$.
 More generally one has to evaluate $X_0(2k_F,\pi,\omega=0)$ to obtain $V$ {\it via}
Eq.~(\ref{eq:rpa}). We first plot the results in Fig.\ref{fig:lindhard}(b).  The physical picture is then as follows: begin at $V=0$ where the system is a quasi-1D Fermi gas.  Start increasing $V$, and the interchain hopping will renormalize to its value $t_\perp^*$.  As $V$ is increased further, eventually the CDW instability Eq. (\ref{eq:rpa}) will be satisfied, with $X_0$ evaluated using the renormalized hopping $t_\perp^*$.  At this point, the system becomes crystalline, and will remain crystalline as V is further increased arbitrarily.  If this occurs after the meta-nematic transition, then both phases will be seen.  If the crystallization happens first (as is the case when $t_\perp=0$), then the meta-nematic transition will not be seen.  This allows the full phase diagram of the Hamiltonian (\ref{eq:H}) to be plotted in Fig. \ref{fig:phase_diagram}.

\section{Smectic phase}

So far, we have been using the nearest neighbor approximation for the interaction  - Hamiltonian \ref{eq:H}.  In this case, the Fourier transform of the interaction potential $V(\mathbf{k})$ is independent of
$k_\|$, hence the dominant CDW instability is always at the peak of the Lindhard
function, i.e. $k_\|=2k_F$.  However, we now consider the full structure of the dipole
interaction \ref{dipole_interaction}.  In particular, when the ratio of lattice spacings $\alpha= a_{\|}/a_{\perp}$ is not large, $V(\mathbf{k})$ acquires a large dependence on $k_\|$.  For small values of the parameter $\alpha$, the largest negative value of $V(\mathbf{k})$ is at $\mathbf{k}=(0,\pi)$.  Although the Lindhard function itself is small at this wavevector (Fig. \ref{fig:lindhard}), the product $V(\mathbf{k})X_0(\mathbf{k})$ may not be, and can become larger than the product $VX_0$ at $(2k_F,\pi)$.  
Hence so long as $t_\perp^* \ne 0$ (i.e. $X_0(k_\|=2k_F)$ is finite), then there is a level of anisotropy of the lattice where the leading instability is at $(0,\pi)$ and not $(2k_F,\pi)$.

\begin{figure}[bt]
\centerline{\psfig{file=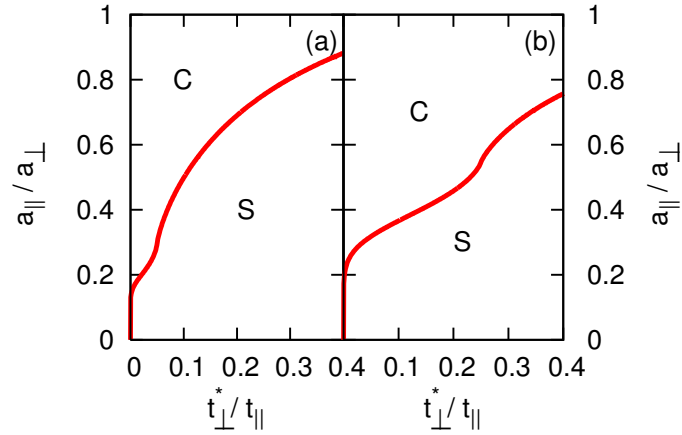,width=3.65in}}
\vspace*{8pt}
\caption{The phase diagram that shows whether the dominant instability is towards the crystalline (C) or smectic phases (S). The axes are the lattice anisotropy $a_\|/a_\perp$ and the relative strength of the perpendicular hopping $t_\perp^*$, for $\mu^* = \pm 1.9 t_{\|}$ (a) and $\pm 1.5 t_{\|}$ (b).}\label{fig:smectic}
\end{figure}

The $(0,\pi)$ instability is still a form of CDW - however as it breaks
lattice symmetry in one direction only, it has smectic order.  Fig.~\ref{fig:smectic} shows which of the two instabilities takes place first as the strength of the interaction is increased. The smectic order is favored when the interaction between neighboring chains is such that the fermions can lower their energy by crowding every other chain, paying a penalty in kinetic energy but lowering the interaction energy. Note that the strong-coupling limit ground state is always a density wave. 

\section{Bosonization Approach}

In the particular case $t_\perp = 0$, the 'bare' dispersion relation is
strictly one-dimensional, allowing us, in principle, to employ the
bosonization technique. In fact, the bosonized Hamiltonian features
backscattering terms that are responsible for the checkerboard
crystallization which, in this limit, takes place at arbitrarily small
coupling \cite{Carr-Tsvelik}. Thus strictly speaking a bosonized description of the fluid of
fermions in this model is never valid. Nevertheless it is illustrative to
ignore the backscattering terms and see what happens. The Hamiltonian can
then be diagonalized exactly to compute the holon dispersion relation in the
abscence of crystallization. 

To bosonize the interacting part of the Hamiltonian first we do a Fourier transformation of all the creation/annilation operators and then we manipulate the order of the operators further to produce a form where only the fermionic density appears.
Note that we will need to work in two-dimensional space since the interactions couple the chains. At the end the interacting part of the Hamiltonian becomes:
\begin{equation}
  \hat{H}_{\mbox{int}}
  =
  -2 V \sum_{\mathbf{q}, q_x > 0} \cos(q_x) \hat{\rho}(\mathbf{q} = 0) 
    + 2 V  \sum_{\mathbf{q}, q_x> 0} \cos(q_x) \hat{\rho}(\mathbf{q})
  \hat{\rho}(-\mathbf{q})    
  \label{eq:H_int2}
\end{equation}
where $V$ is the potential and the direction $x$ is the one perpendicular to the chains. The first term then shifts the chemical potential since $\hat{\rho}(0)$ is the density operator at ${ \mathbf{q}}=0$ (in other words
it counts the number of fermions).

The most interesting four-fermion term then, provides in the language of Ref.~\cite{Schulz} : $g'_2 = g'_4 = 2 V \cos(q_x)/ (2 \pi v_F)$
where $v_F$ is the Fermi velocity.
Then the dispersion relation becomes $u_{\rho}(\mathbf{q}) |q_y|$, with:
\begin{eqnarray}
  u_{\rho}(\mathbf{q}) 
  &=& 
  v_F 
  \sqrt{
    \left[
      1 +  \frac{2V\cos(q_x)}{\pi v_F} 
    \right]^2  
  -
    \left[
      \frac{2V\cos(q_x)}{ \pi v_F}
    \right]^2 
  }
\end{eqnarray}

If the expression under the square-root becomes negative then the energy of low-lying excitations becomes unphysically imaginary, signalling the breakdown of bosonization. The condition for this to take place is:
 $v_F > - 4 V \cos(q_x)/ \pi$ which, if it holds for every $q_x$, demands $\pi/4 v_F >  V$.
Thus we find that the holon velocity vanishes at a
critical coupling which corresponds to the onset of the smectic phase and
coincides exactly with a divergence of the RPA susceptibility at ($\pi$,0) at
the same critical coupling $V_{crit}$. Beyond $V_{crit}$, the bosonized theory has
broken down and it is no longer useful. Moreover we stress that even below
this value the use of bosonization is somewhat artificial, since we need to
cross out the backscattering terms that, in reality, will lead to the
checkerboard crystallisation before any of the physics that we have
discussed on the basis of that approach take place (as our RPA approach
shows). However it affords a different perspective on the nature of the
transition into the smectic state. In particular, it shows that the smectic
instability can be interpreted in terms of plasmons going soft as we
approach it. Moreover, it lends further credibility to our RPA approach,
which can be applied over the whole phase diagram and used to find both the
instabilities into the smectic and crystalline states, as the critical
coupling $V_{crit}$ obtained with bosonization for the smectic phase coincides
with that given by bosonization, in the appropriate limit $t_\perp  \rightarrow 0$
(although, we stress once again, in that limit the crystallisation takes
place first).
 
%Recalling that $v_F =  d\mathcal{E}_{k}/dk|_{k=k_F}$ ($\hbar = 1$ everywhere)
%and that 
%$D(\mathcal{E}) = 2/\pi ~ dk/d\mathcal{E}_{k}$, where $\mathcal{E}_k$ is the 1D dispersion relation, then the density of states at the Fermi energy is $D(0) = 2/(\pi v_F)$ and the condition we obtain coincides with a Stoner-like %criterion we will obtain in the next paragraph through Random Phase Approximation (for $t_\perp=0$), which leads to a crystalline phase. The velocity in the bosonization approach is the {\em plasmon} velocity, and the low energy %propagating elementary excitations are still plasmons, and not Fermions.

\section{Conclusions}

With the help of optical lattices and cold atoms we can realise correlated fermion physics.
In particular interactions between dipolar  atoms or molecules can lead to quantum soft-matter phases, hypothesized in strongly interacting quantum systems.
Meta-nematic, smectic, and crystalline phases are predicted to occur for fermions.
For bosons the expected phases are either a superfluid phase or a CDW.

The tunability of the experimental parameters in optical lattices leads to
many possibilities for experiments with polarised fermions.
For example, by orienting the field along the chain direction, we could
realise a model with attractive intra-chain interactions and repulsive
inter-chain interactions, presumably leading to superconducting\
 stripes.
Using the same set-up discussed here, but with the chains perpendicular to the
direction with no interactions, we can have one-dimensional Luttinger liquids
and then, by adding transverse hopping one can study the fundamental confinement-deconfinement transition and orthogonality
catastrophe.

Another possible extension of the present study is the three-dimensional
optical lattices as well as the regime of finite temperatures.

\section*{Acknowledgements}

JQ gratefully acknowledges an Atlas Research Fellowship awarded by CCLRC (now STFC) in association with St. Catherine's College, Oxford.
STC was funded by EPSRC grant GLGL RRAH 11382.

\end{document}